\documentclass[utf8]{frontiersFPHY} 
\pdfoutput=1
\setcitestyle{square} 
\usepackage{url,lineno,microtype,subcaption}
\usepackage[onehalfspacing]{setspace}
\usepackage{tikz}
\usepackage{import}
\usepackage{float}
\usepackage{booktabs}

\def\keyFont{\fontsize{8}{11}\helveticabold}
\def\firstAuthorLast{Kumar {et~al.}} 
\def\Authors{Sandeep Kumar\,$^{1,*}$, Arghyadeep Paul\,$^{1}$ and Bhargav Vaidya\,$^{1,2}$}
\def\Address{$^{1}$ Indian Institute of Technology Indore, Discipline of Astronomy Astrophysics and Space Engineering, Indore, 453552, Madhya Pradesh, India \\
$^{2}$ Indian Institute of Science Education and Research Kolkata, Centre of Excellence of Space Science in India (CESSI), Mohanpur, 741246,  West Bengal, India}
\def\corrAuthor{Corresponding Author: Sandeep Kumar}
\def\corrEmail{ssandeepsharma201@gmail.com}

\begin{document}
\begin{nolinenumbers}
\onecolumn
\firstpage{1}

\title[Forecasting Solar Wind]{A comparison study of extrapolation models and empirical relations in forecasting solar wind} 

\author[\firstAuthorLast ]{\Authors} 
\address{} 
\correspondance{} 

\extraAuth{}

\maketitle

\begin{abstract}
\section{}
Coronal mass ejections (CMEs) and high speed solar streams serve as perturbations to the background solar wind that have major implications in space weather dynamics. Therefore, a robust framework for accurate predictions of the background wind properties is a fundamental step towards the development of any space weather prediction toolbox. In this pilot study, we focus on the implementation and comparison of various models that are critical for a steady state, solar wind forecasting framework. Specifically, we perform case studies on Carrington rotations 2053, 2082 and 2104, and compare the performance of magnetic field extrapolation models in conjunction with velocity empirical formulations to predict solar wind properties at Lagrangian point L1. 
Two different models to extrapolate the solar wind from the coronal domain to the inner-heliospheric domain are presented, namely, (a)  Kinematics based (Heliospheric Upwind eXtrapolation [HUX]) model and (b) Physics based model. 
The physics based model solves a set of conservative equations of hydrodynamics using the PLUTO code and can additionally predict the thermal properties of solar wind. 
The assessment in predicting solar wind parameters of the different models is quantified through statistical measures. We further extend this developed framework to also assess the polarity of inter-planetary magnetic field at L1. Our best models for the case of CR2053 gives a very high correlation coefficient ($\sim$ 0.73-0.81) and has an root mean square error of ($\sim$ 75-90 kms$^{-1}$). Additionally, the physics based model has a standard deviation comparable with that obtained from the hourly OMNI solar wind data and also produces a considerable match with observed solar wind proton temperatures measured at L1 from the same database.

\tiny
 \keyFont{ \section{Keywords:} Solar Wind, Sun-Earth connection, Sun: Magnetic fields, Sun: Heliosphere, Method: Numerical, Space Weather Modelling} 
\end{abstract}

\section{Introduction}
Space weather refers to the dynamic conditions on the Sun and in the intervening Sun-Earth medium that can severely influence the functioning of space-borne and ground based technical instruments thereby affecting human life. 
Predicting the impact of space weather thereby becomes an essential task. In particular, explosive events on the Sun that include solar flares, coronal mass ejections (CMEs) and solar energetic particles (SEPs) play a crucial role in influencing space weather \cite{Schwenn:2006}. 
The ambient solar wind being the medium in which the CMEs propagate also plays a significant role in influencing space weather, particularly, high speed solar wind streams which contribute to about 70\% of geomagnetic activity outside of the solar maximum phase \cite{Richardson:2000}. 
Therefore, understanding and predicting the key properties of the ambient solar wind is a crucial component of space weather modelling \cite{Owens:2013}.

Magnetic field is a key ingredient that threads the solar plasma and governs the dynamical properties of solar wind. 
Observational measurements of field strengths in tenuous coronal environments is a challenging task even today. 
Thus, modelling coronal plasma requires extrapolating magnetic fields at the photosphere in the coronal region. 
Typically, a potential field source surface solution (PFSS,  \cite{Altschuler:1969}) is adopted to extend photospheric magnetic fields up-to the source surface, usually set at $\sim 2.5 R_{\odot}$). 
Further, solutions from PFSS is augmented with magnetic fields obtained from the Schatten current sheet model (SCS, \cite{Schatten:1972}). 
This ensures confining heliospheric currents into a very thin sheet in accordance to \textit{Ulysees} measurements of latitude independent radial interplanetary field component \cite{Wang:1995}. 
Field extrapolation techniques present an alternative approach to more computationally demanding magneto-hydrodynamic (MHD) simulations for estimating coronal properties from input photospheric magnetic field data (\cite{Lionello:2009} and references therein). 
On comparing the properties of open magnetic field lines at the source surface with observed solar wind velocity, empirical relations have been formulated viz., Wang-Sheeley model \cite{Wang:1990} and its improvement Wang-Sheeley-Arge model \citep{Arge:2000, Arge:2003} to calculate velocity at that surface. 
A state-of-the-art solar wind forecasting framework combines the coronal model described above with the inner-heliospheric model to estimate wind parameters at L1. Kinematic extrapolation methods that rely on 1D stream propagation like Heliospheric Upwind eXtrapolation \cite{Reiss:2019} and its time dependent variant HUXt \cite{Owens:2020} provides a computationally efficient solution without providing physical insight as done by the 3D MHD physics based models like ENLIL \cite{Odstrcil:2004}, SWMF \cite{SWMF_newer}, EUHFORIA \cite{Pomoell:2018}.  
The different approaches adopted for forecasting solar wind along with their quality assessment are presented in a review by MacNeice et. al. \cite{MacNeice:2018}.  

The various forecasting models over time have shown considerable progress in our understanding of modelling the macro-physical solar wind properties. 
One of the key ingredients that is critical for space weather modelling is the interplay of micro-physical turbulent and particle acceleration processes with macro-physical dynamics. 
Several attempts have been made to include effects of energy contained in sub-grid turbulence and multi-fluid aspects in modelling of solar wind (CORHEL \cite{Downs:2016}, AWSoM\cite{AWSOM}, CRONOS \cite{Weingarten:2016}, \cite{Usmanov:2014}, \cite{Usmanov:2018}).
In general, handling of large separation of scales required to consistently model such an interplay of length scales is a challenging task. 
Such a multi-scale nature of problem is also experienced in astrophysical plasma modelling and several astrophysical codes are working in direction of developing a hybrid framework that bridges this wide gap. 
In particular, recent development of hybrid particle module for PLUTO code \citep{Mignone:2018,Vaidya:2018} to model particle acceleration at shocks and its subsequent non-thermal emission. 
Further, PLUTO code supports adaptive mesh refinement and various non-ideal MHD processes including magnetic resistivity \cite{Mignone:2012} and Hall-MHD. The code also has support for anisotropic thermal conduction \cite{Vaidya:2017} and optical thin cooling. 
In last 5 years, problems pertaining to solar and magnetospheric physics have also been tackled with PLUTO code \citep{Reale:2016, Sarkar:2017, Das:2019, Reville:2020}. 
The goal of this work is to develop a space weather modelling framework in conjunction with PLUTO code aiming to utilize the additional functionalities of the code for modelling the micro-physical aspects of Sun-Earth environment. 

In this first study, our focus is to compare the various coronal and inner heliospheric models for solar wind forecasting and the assessment of their predictive performance. 
The paper is arranged in the following manner - the details of the methods implemented for the forecasting framework are described in Sec~\ref{sec:methods}. 
The results from various models adopted for forecasting of solar wind velocity for a couple of case studies along with statistical assessment of their accuracy are elaborated in Sec~\ref{sec:results}. 
Finally, the Sec~\ref{sec:discuss} discusses the various features of the forecasting framework along with limitations and future outlook. 

\section{Methodology}
\label{sec:methods}
The region between the solar photosphere and the Lagrangian point L1 is divided into two zones.  
The inner coronal zone extends from the photosphere up-to 5 R$_{\odot}$, followed by inner-heliosphere zone that extends from 5 R$_{\odot}$ up-to L1 point. 
Data driven prediction of solar wind parameters at L1 point requires the following steps - 
\begin{itemize}

\item To calculate the magnetic fields in the coronal region through various extrapolation methods of the input observed photospheric magnetic field.
\item Applying the velocity empirical relations based on the field line properties obtained from extrapolation at the outer boundary of the coronal region. 
\item Extending the velocity estimates from the outer boundary of the coronal region up-to Lagrangian point L1 for comparison with observations. 
\end{itemize}

Detailed procedure followed for each of the above steps is described in this section.

\subsection{Magnetic Field Extrapolation}
Forecasting of solar wind across the domain requires accurate magnetic field solutions extrapolated from the solar surface to the outer boundaries of the coronal domain. The magnetic field extrapolations are carried out up to a distance of 5$R_\odot$. The inner boundary conditions for the magnetic fields at the solar surface are specified by the input magnetograms(synoptic maps) taken from the Global Oscillations Network Group(GONG) (\url{https://gong.nso.edu/data/magmap/crmap.html}). Once the inner boundary conditions are specified, the extrapolation of the magnetic fields are then carried out in a twofold manner-
Just above the photospheric region, the magnetic energy density is greater than the plasma energy density and magnetic effects dominate. One can assume this region to be current free and thus, use the potential formulation for the magnetic fields\cite{sch_w_1968}. This current free assumption is valid inside a sphere of radius of about 2.5$R_\odot$, the outer boundary of which is known as the \textit{source surface}. The magnetic fields inside the source surface can be solved using the PFSS (Potential Field Source Surface) model. The outer boundary condition for this model dictates that the magnetic fields at the source surface is approximately radial\cite{sch_w_1968}. For the PFSS solution, we used the python module \textit{pfsspy}, which is an open source, finite difference PFSS solver. Using the observed magnetogram data as an input, potential field equations are solved in radial direction on a logarithmic grid, whereas for latitude and longitude the grid is regularly spaced in terms of $\cos\theta$ (-1.0 to 1.0) and $\phi$ (0 to 2$\pi$) \citep{david_stansby_2019_2566462,anthony_yeates_2018_1472183}. Field line tracing using the field solutions is done via Runge Kutta 4$^{th}$ order method.
In the region outside the source surface, the magnetic fields are extrapolated using the Schatten Current Sheet (SCS) model \cite{Schatten:1972}. 
The SCS model extends the magnetic fields from the source surface to a distance of 5$R_\odot$, i.e, the outer boundary of the coronal zone. 
The input to the SCS model is the re-oriented output from \textit{pfsspy}, i.e, if at the source surface, $B_r\geqslant 0$, the field remains unchanged, but if $B_r\leqslant 0$, $B_r, B_\theta, B_\phi$ are replaced by $-B_r, -B_\theta, -B_\phi$.
Using the same resolution in $\cos\theta$ and $\phi$ plane as that of the output from \textit{pfsspy}, all magnetic field components 
beyond the source surface can be expressed in the form of a Legendre polynomial expansion \cite{Schatten:1972,Reiss:2019}. 
Following the formulations presented in the appendix of \cite{Reiss:2019}, we construct the matrices of coefficient of spherical harmonics $g^{m}_{n}$ and $h^{m}_{n}$, where 
$n$ and $m$ are degree and order of the associated Legendre polynomial $P^{m}_{n} (\cos\theta)$. A rather modest value of n = 15 is used to approximate the magnetic field values using the SCS model. The accuracy of SCS approximation improved only marginally with doubling the choice of $n$, but the computational time increased significantly. Thus, the order of Legendre polynomial used in the present work was carried out by optimising both accuracy and computational time.
While tracing the field lines, one should note that the boundary conditions of PFSS and SCS are not compatible with each other. A direct combination of the PFSS solution with the SCS model at 2.5 $R_\odot$ results in kinks and discontinuities at the model boundary which is a consequence of the different boundary requirements for the two models. To avoid this non-physical discrepancy, the input to the SCS model is given by the field values at 2.3 $R_\odot$ rather than 2.5$R_\odot$. The field values in the thin radial slice between  2.3 $R_\odot$ and 2.5 $R_\odot$ is then overwritten by the values obtained by the SCS model. This leads to a smooth transition and a more desirable coupling between the PFSS and the SCS models \cite{mc_2008}. The SCS model then extrapolates the fields upto 5 $R_\odot$. The PFSS + SCS solution together gives us a good approximation of the magnetic field structure up-to a distance of 5 $R_\odot$.

\subsection{Velocity Empirical Relations}
Based on the magnetic field structure\cite{Reiss:2019} obtained by from the field extrapolation methods, we generate a velocity map for the solar wind using some empirical velocity mapping models. We employ two different empirical models for the velocity calculations described in the sections below.

\subsubsection{Wang-Sheeley Model(WS)}
The Wang-Sheeley model depends on a parameter $f_s$, called the \textit{expansion factor} of the coronal flux tubes to calculate Solar Wind velocities. The expansion factor($f_s$) is given by
\begin{equation}
    f_s= (R_\odot/R_{ss})^2 [B^P (R_\odot)/B^P (R_{ss})] \label{fs}
\end{equation}
where $B^P(R_{ss})$ denotes the radial field strength at a sub-earth point \textit{P} on the source surface and  $B^P (R_\odot)$ is the foot-point of the flux tube traversing \textit{P} on the photosphere\cite{Wang:1990}. The \textit{expansion factor} measures the amount of change in cross section of a flux tube between the photosphere and the source surface compared to a purely radial expansion. It is observed that there is a correlation between the expansion factor and the solar wind velocities. Based on this, an empirical formula for calculating the solar wind velocities can be devised -
\begin{equation}
    v_{sw}^{ws} (f_s)= v_{\rm slow} + \frac{v_{\rm fast}-v_{\rm slow}}{f_s^{\alpha}} 
\end{equation}
where, $v_{\rm slow}$ is the lowest expected speed as $f_s \to \infty$ and $v_{\rm fast}$ is the fastest solar wind expected as $f_s \to 1$.
For our calculations, we have used the values of $v_{\rm slow}= 200$\,kms$^{-1}$, $v_{\rm fast}= 750\,$ kms$^{-1}$ and $\alpha= 0.5$. \cite{Arge:2000}.

\subsubsection{Wang-Sheeley-Arge Model(WSA)}
The Wang-Sheeley-Arge\cite{wsa_2003} is a model that also incorporates the effect of minimum angular distance of foot point of the field line from coronal hole, along with the expansion factor. It is believed that coronal holes produce fast streams of solar wind. So the position of the field line's foot point in coronal hole plays a very important role. 
The empirical relation from the WSA model\cite{Riley:2015} used in the present work is - 
\begin{equation}
    v_{sw}^{wsa} (f_s, \theta_b)= v_{\rm slow} + \frac{v_{\rm fast}-v_{\rm slow}}{(1+ f_s)^{\alpha}} \left( 1 - 0.8 e^{-(\theta_b /w)^{\delta}} \right)^{3.5} 
    \label{eq:wsa}
\end{equation}
The parameters $v_{\rm slow}$ and $v_{\rm fast}$ corresponds to the velocity of fastest and slowest solar wind stream. $\theta_b$ is the minimum angular distance for the foot point of the field line from a coronal hole boundary at the Solar surface. In the present work, we have used $v_{\rm slow}=250$ kms$^{-1}$, $v_{\rm fast}=900$\,kms$^{-1}$, 
$\alpha=1.5/9$, $w=0.03$, and $\delta=1.5$.\\

\subsection{Extrapolation into the Heliospheric domain}
The WS and WSA model gives us the solar wind maps at the outer boundary of the coronal domain i.e., at 2.5 $R_\odot$ or 5 $R_\odot$ based on the choice of magnetic field extrapolation method. 
For comparison with the observed velocities at L1 point, it is required to extrapolate these velocities into the inner heliosphere zone. This requires coupling of the coronal velocity models with heliosperic velocity extrapolation models. 
We have employed two such extrapolation methods, \textit{(a) Heliospheric Upwind eXtrapolation(HUX)} \cite{Riley:2011} and \textit{(b) Physics based Modelling using PLUTO code}. We describe these methods in the following sections.

\subsubsection{Heliospheric Upwind eXtrapolation(HUX)}
The HUX model assumes the solar-wind flow at the outer boundary of the coronal domain to be time-stationary. The extrapolation of the solar wind velocities in an $r$-$\phi$ grid can be then be kinetically approximated using
\begin{equation}
v_{r+1,\phi}=v_{r,\phi}+\frac{\Delta r\Omega_{\odot}}{v_{r,\phi}}\left(\frac{v_{r,\phi+1}-v_{r,\phi}}{\Delta\phi}\right) \label{eq:hux}  \end{equation}\\
where, $\Delta r=1R_\odot$ and $\Delta \phi=1^\circ$ represent the grid spacings in $r$ and $\phi$ directions respectively. $\Omega_{\odot}$ is the angular velocity of the Sun calculated assuming a rotation time period of 27.3 days. 
The HUX is essentially a 1D extrapolation that neglect the effects of magnetic fields, pressure gradients and gravity. The advantage being that such an extrapolation method is computationally inexpensive when compared to state-of-the-art 3D MHD models.

\subsubsection{Physics based Modelling using PLUTO code}
\label{sec:pluto_init}
With an aim to incorporate the effects of some of the physical aspects in solar wind extrapolation, we describe a physics based modelling approach that involves solving a set of conservative equations using the Godunov scheme based Eulerian grid code, PLUTO \cite{Mignone:2007}. 

For this pilot study, we assume the solar wind to be hydrodynamic and solve the following set of compressible equations in 2D polar co-ordinates (r, $\phi$) : 
\begin{eqnarray}
\frac{\partial \rho}{\partial t} + \nabla \cdot ({\rho \vec{v}}) &=& 0 \\ \nolinenumbers
\frac{\partial \rho \vec{v}}{\partial t} + \nabla \cdot ({\rho \vec{v}\vec{v} + P\mathcal{I}}) &=& \rho \vec{g} \\\nolinenumbers 
\frac{\partial E}{\partial t} + \nabla \cdot ((E + P)\vec{v}) &=& \rho \vec{v}\cdot\vec{g}  \nolinenumbers
\end{eqnarray}
where, $\rho$ is the density of the fluid, $P$ being the isotropic thermal pressure, $\vec{v}$ is the fluid velocity, $\mathcal{I}$ is an identity matrix and the total energy $E = \frac{1}{2}\rho v^{2} + \rho \epsilon$ is sum of the kinetic energy and the internal energy. The acceleration due to gravity $\vec{g} = -GM_{\odot}/r^2$ is included as a source term in the conservative momentum equation. A poly-tropic equation of state is adopted with
the value of adiabatic index $\gamma = 1.5$ \citep{Odstrcil:2004, Pomoell:2018}. 

The above equations are solved in a non-inertial frame where the inner radial boundary rotates with the rate equal to the solar rotation rate. Further to simplify we neglect the Coriolis and centrifugal terms as their contribution is rather small in determining the steady flow structure of the solar wind \cite{Pomoell:2018}. 
The computational grid in polar co-ordinates ranges in the radial direction from 5$R_{\odot}$ to 435$R_{\odot}$ with a resolution of 512 grid cells. Same number of grid cells are used to resolve the azimuthal ($\phi$) direction. 

Initially, the computational domain is filled with a static gas with number density of  1 cm$^{-3}$ and thermal sound speed of 180 km s$^{-1}$. The choice of initial conditions does not affect final steady state wind solution as the material injected from the inner boundary wash away these initial values to obtain a new steady state.
Radial velocity is prescribed at the inner boundary from the WSA mapping. The inner boundary is also made to rotate with respect to the computational domain with a rate equal to the solar rotational time period which is 27.3 days. 
As a general rule, boundary conditions can be specified for only those characteristics that are outgoing (away from the boundary). 
Since we have a supersonic inflow boundary condition, all the characteristics are pointing away from the boundary and therefore, along with the prescription of solar wind velocity, one would need to prescribe the density and pressure at the inner radial boundary as well. 
Following \cite{Pomoell:2018}, we prescribe the number density (n) and pressure at the inner radial boundary in following manner : 
\begin{eqnarray}
    n(r) &=& n_{0} \left(\frac{v_{0}}{v_{r}}\right)^2\\\nolinenumbers
    P &=& P_0
\end{eqnarray}
where $P_0$ is set to be a constant value of 2.75 nPa and $n_{0} = 100$ cm$^{-3}$ resulting in a proton temperature of 2 MK. The scaling velocity $v_{0}$ is set to be 675 km s$^{-1}$. The outer radial boundary is set to have free flowing outflow conditions. 
For each Carrington rotation, we carry out the simulations using velocities obtained from sub-earth points with the magnetic field parameters obtained at 5 R$_{\odot}$ and employing WSA empirical relation for velocity. The results are presented in Sec~\ref{sec:fc_pluto}.

\subsection{Model Definitions}
Using the combination of various magnetic field extrapolation, velocity empirical relations and method of extrapolating velocity field within the inner heliosphere, we have defined 6 different models. 
The combinations used for each of these models are shown in Figure ~\ref{fig:flowchart}. 
Models 1 and 2 involves magnetic field extrapolation using PFSS (without SCS) and velocity extrapolation in inner heliosphere using the HUX model. 
The difference between them is the choice of velocity empirical relation. 
In the other set of models 3 and 4a, the magnetic fields are extrapolated using PFSS+SCS. 
The empirical velocity relation employed for Model 3 is WS and that for Model 4a is WSA. Velocity field in both these models are extrapolated from 5 R$_{\odot}$ to L1 using HUX. 
Model 4b is a variant obtained by considering an ensemble values using the same combinations as that of Model 4a (see section~\ref{sec:ensemble_fc}). Model 4c uses the PFSS+SCS field extrapolation model and WSA velocity empirical relation similar to that used by Model 4a, however, the extrapolation of velocity field is done using physics based hydrodynamic simulations. 

\begin{figure}
\centering
\includegraphics[width=\textwidth]{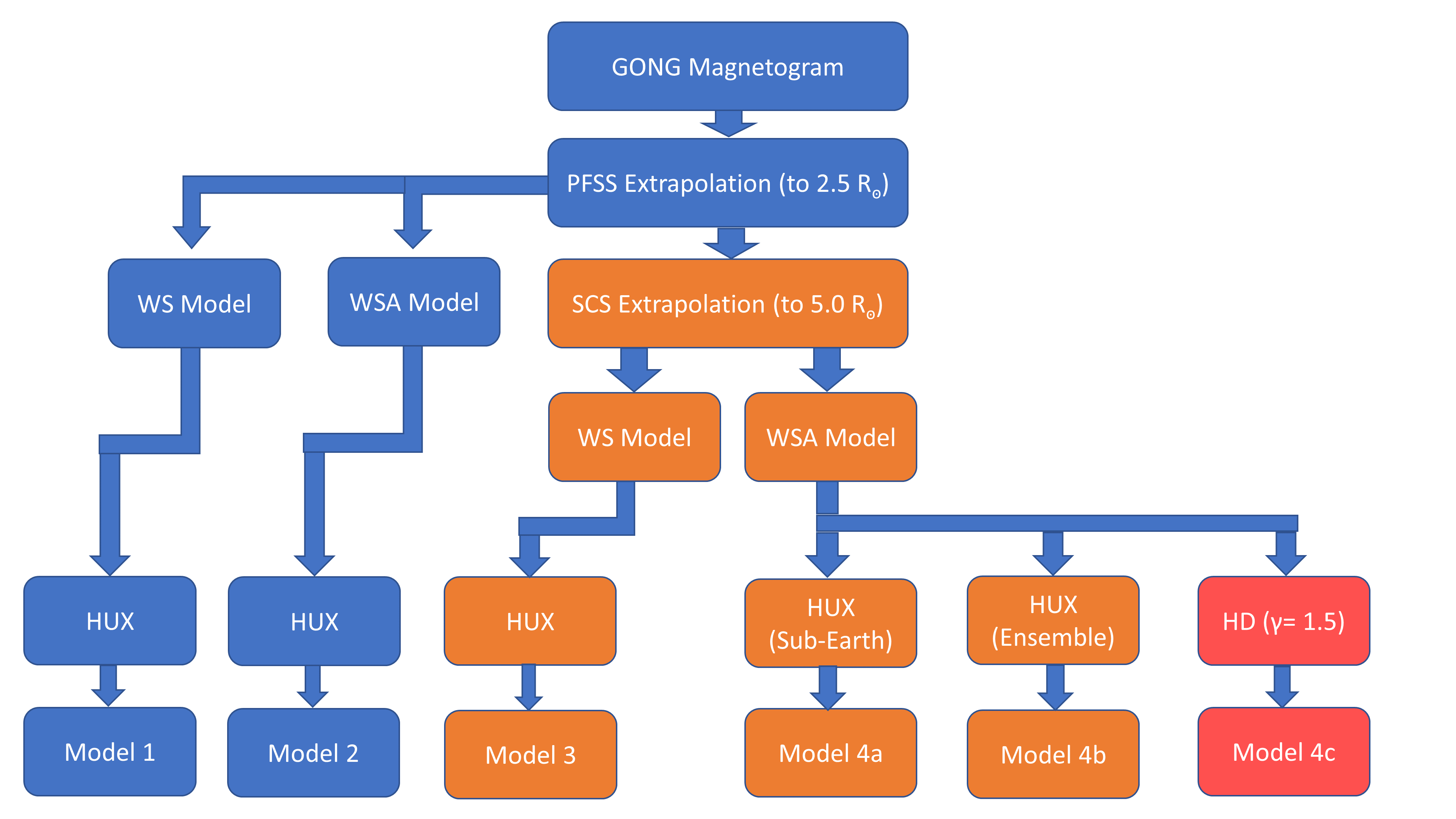}
\caption{A flowchart of the models that have been utilized in the present work. The various combinations of coronal and inner heliospheric models have been categorised into six different models : 1, 2, 3, 4a, 4b and 4c.}
\label{fig:flowchart}
\end{figure}

\subsection{Statistical Measures of Forecast Performance}
\label{sec:stats}
The performance of a forecast can be determined by comparing the forecast outcome of continuous variables (e.g., velocity) to the observed values. We calculated several scalar measures of forecast accuracy which has previously been used to determine forecast performances by \citep{Reiss:2016, Wu:2020aa}. 
Given a set of modelled values $m_n$ and a set of corresponding observed values $o_n$,  the Mean Absolute Error (MAE) is given as the arithmetic mean of the absolute values of the differences between the model output and the observed values at each observed data point. 
\begin{equation}
    MAE= \frac{1}{N} \sum_{n=1}^N \mid m_{n} - o_{n} \mid
\end{equation}

The Mean absolute error can be considered as a measure of the overall error in forecast of a model. Another such measure is the Mean Absolute Percentage error(MAPE) which is given by 
\begin{equation}
    MAPE= \frac{100}{N} \sum_{n=1}^N  \left\lvert \left(\frac{m_{n} - o_{n}}{o_n}\right) \right\rvert
\end{equation}
The Root Mean Square Error or RMSE is also used sometimes as a performance statistic for a model and is given by 
\begin{equation}
    RMSE=\sqrt{ \frac{1}{N} \sum_{n=1}^N  \left(m_{n} - o_{n}\right)^2}
\end{equation}

Another important measure in determining the forecast performance is the Pearson Correlation Coefficient (CC) which is a parameter used to estimate the correlation between the observed values and the model values. In addition to this, another measure is the standard deviation of the time series of each of the individual model outputs and its comparison with the estimate from observed data.  
These measures are estimated for all the models considered and discussed in section ~\ref{sec:stats_res} for the three Carrington rotations CR2053, CR2082 and CR2104 considered in our study.
\section{Results}
\label{sec:results}
In this section, we present our results using the methodology described above for the three case studies spanning the declining phase of cycle 23, near the solar minimum and the rising phase of solar cycle 24. 
In particular, we consider CR 2053 (from 2007/02/04 to 2007/03/04), CR 2082 (from 2009/04/05 to 2009/05/03) and CR 2104 (from 2010/11/26 to 2010/12/24).We also demonstrate the assessment of performance of the different forecasting models considered for these cases. 
\subsection{Case Studies}
CR2053 represents a relatively quiet phase of the sun during the decline of the solar cycle number 23. Six active regions were identified regions during the period of CR2053 \cite{2016ApJ} that can also be seen in our input synoptic magnetogram (panel (A) of Fig.~\ref{fig:PFSS_2053}). All the active region lie close to the L1 footprint and thus are expected to have pertinent effects on the solar wind velocities.
\begin{figure}
\centering
\includegraphics[width=\textwidth]{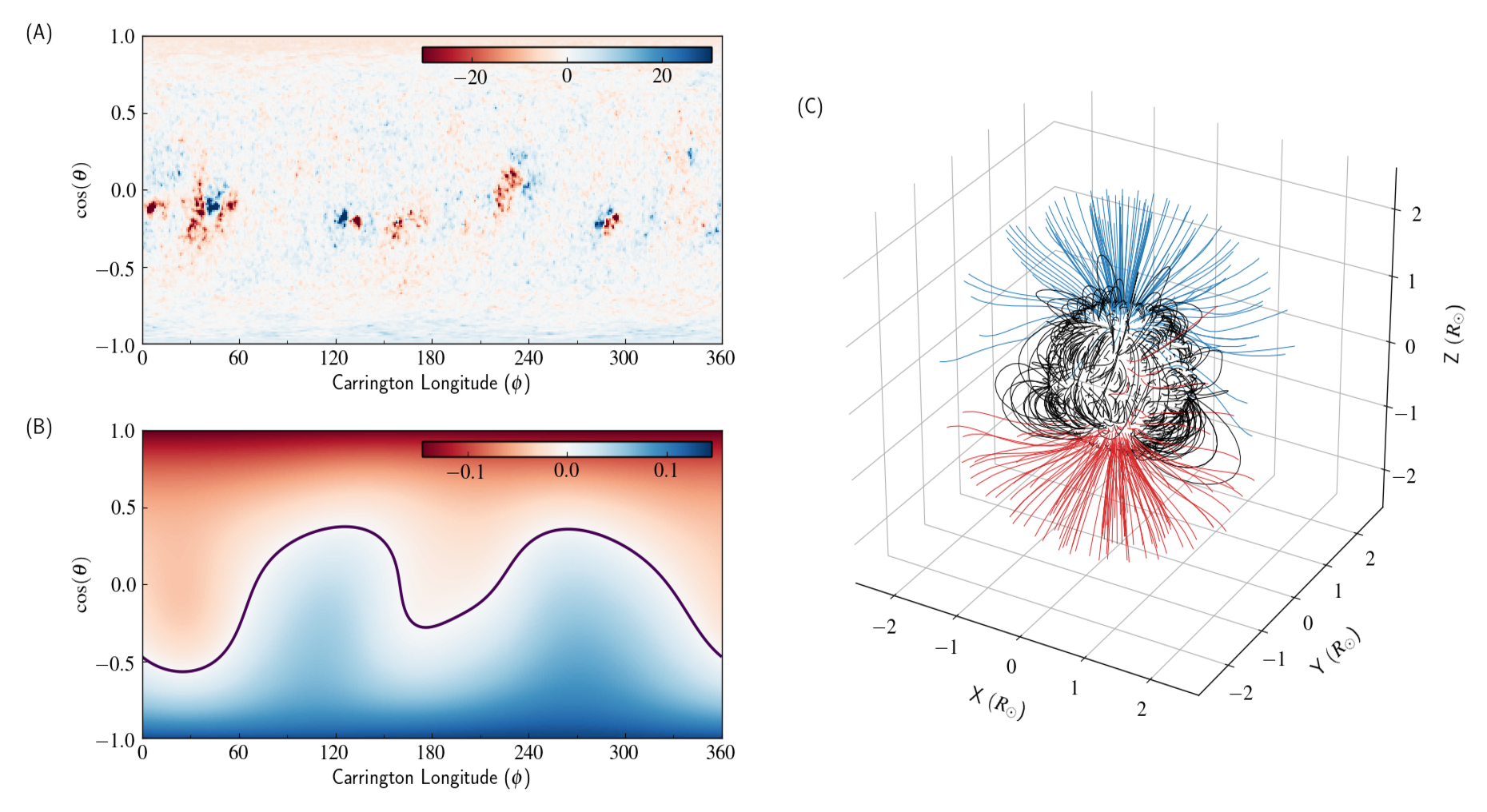}
\caption{Panel (A) shows the photspheric magnetic flux density (in Gauss) from a standard synoptic magnetogram obtained from GONG for CR 2053, panel (B) shows the magnetic flux density measured in Gauss obtained from PFSS extrapolation at the source surface (2.5 R$_{\odot}$) with the polarity inversion line shown in violet. The distribution of magnetic field lines obtained from the PFSS solution is plotted in panel (C). 
The field lines having negative polarity are shown in \textit{blue} and the field lines having positive polarity are shown in \textit{red}. The closed field lines are represented in \textit{black}. The magnetic fields have units in Gauss and the distances on the radial axis are normalised to $R_\odot$}
\label{fig:PFSS_2053}
\end{figure}
\begin{figure}
\centering
\includegraphics[width=\textwidth]{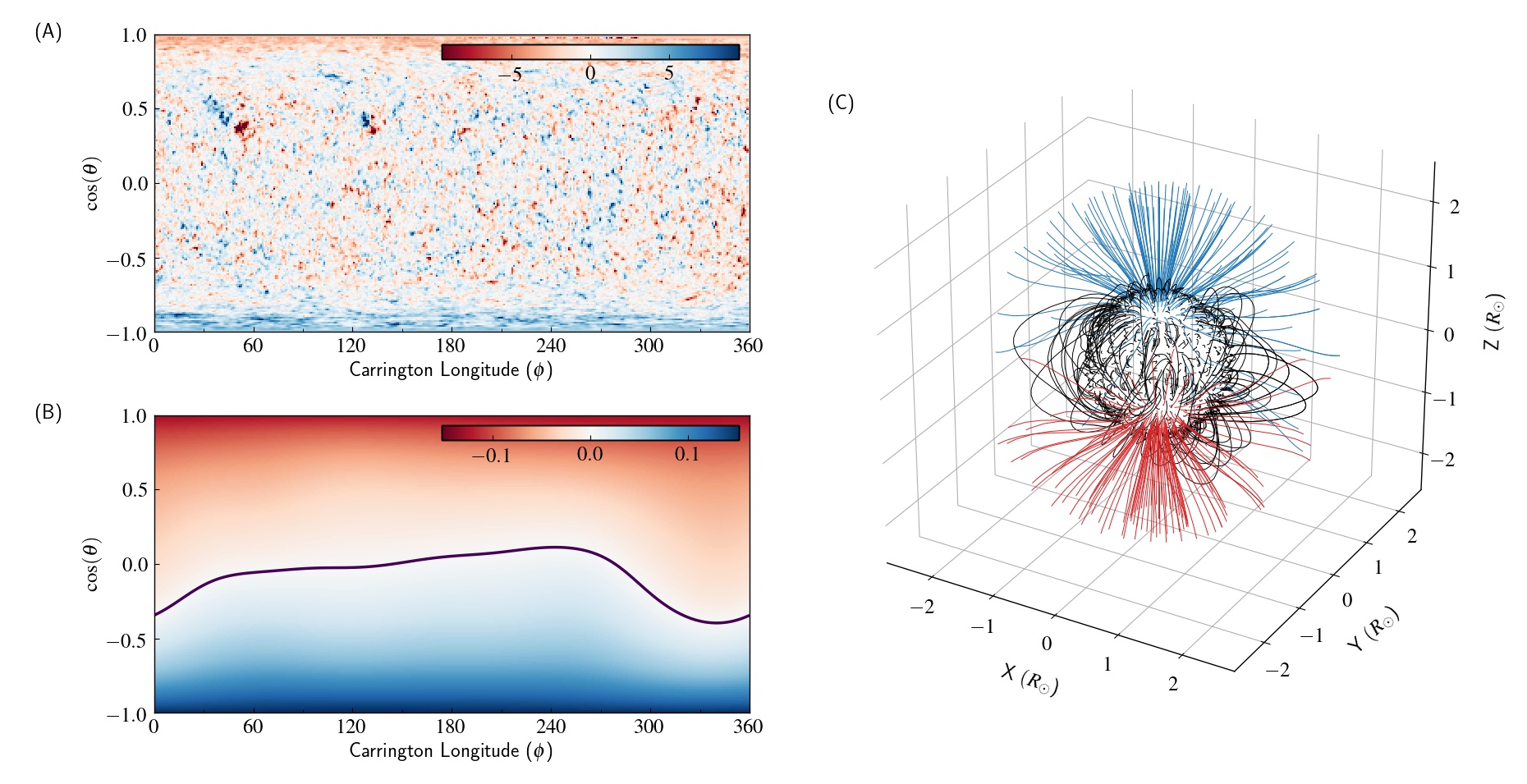} 
\caption{The PFSS input and output in a similar layout as Figure: \ref{fig:PFSS_2053} for CR 2082.} 
\label{fig:PFSS_IO_2082}
\end{figure}
As a first step, we calculated the PFSS solution for this Carrington rotation. 
The GONG magnetogram data which is given as input to PFSS and the output solution obtained using \textit{pfsspy} are shown in Figure.~\ref{fig:PFSS_2053}. 
A 3D map of field lines joining the photosphere to the PFSS source surface at 2.5 R$_{\odot}$ is shown in panel (C) of the same figure. The angle $\theta$ on the Y-axis of panels (A) and (B) is related to Carrington latitude ($\theta_{\rm cr}$) as $\theta = -\theta_{\rm cr} + 90^\circ$.
It is evident from the distribution of field lines that few open field lines with both positive (red) and negative (blue) polarities have their foot points located on the photosphere that are close to solar equator. On extending these field lines from source surface up-to 5 R$_{\odot}$ incorporating the SCS model, we obtain a distribution of magnetic field lines at the boundary of coronal domain. 
Sub-earth field lines are selected from such a distribution and the parameters like the expansion factor ($f_{s}$) and their distance to a nearby coronal hole boundary ($\theta_b$) are estimated. These parameters obtained at 5 R$_{\odot}$ are shown in panels  (A) and (B) of figure ~\ref{fig:fs_2053} as a function of Carrington longitude ($\phi$).
The solar wind velocity obtained from using these parameters in the empirical velocity relations of WS (Model 3) and WSA (Model 4a) are shown in panel (C) in comparison with the observed data. 
As the models 1 and 2 only rely on the magnetic fields interpolated using PFSS solution, the field parameters are estimated at 2.5 R$_{\odot}$ and used in the empirical relations to estimate the velocity at L1 after HUX extrapolation. 

We observe that the WS model gives an inaccurate representation of the solar wind velocities and the contrast between the slow wind and the fast streams are not satisfactorily reproduced. For example, during the first phase of high speed stream peaking on 2007-02-15, variations are observed in values of $\theta_{b}$ of the order of 0.03 radians, while the variation in the expansion factor ($f_{s}$) is not significant. This translates into a change in the estimate of the empirical velocity at 5 R$_{\odot}$. The WS extrapolation that is solely dependent on value of $f_s$ do not show appreciable variation resulting in poor prediction for models 1 and 3. 
Whereas, models 2 and 4a capture the high speed stream due to its dependence on $\theta_{b}$.
The case of CR2082 also illustrates the quiet phase of the Sun, in-fact it lies at the solar minimum at the end of cycle 23. The input GONG magnetogram is shown in panel (A) of figure~\ref{fig:PFSS_IO_2082}. The solution obtained from PFSS at the source surface shows a rather bipolar structure, where the polarity inversion line is relativity straight lying around the co-latitude $\cos(\theta) \sim 0$. The bipolar structure is also evident from the 3D distribution of magnetic field lines connecting thephotosphere with the source surface shown in panel (C) of the same figure. 

Extrapolated field lines from PFSS solution and ones that are augmented with SCS model are used to obtain the values of the expansion factor $f_{s}$ and coronal hole boundary distance $\theta_{b}$ at sub-earth points. 
Panels (A) and (B) of figure ~\ref{fig:fsthb_vel_2082} shows these parameters for the model where field line is extrapolated up-to 5 R$_{\odot}$ using PFSS + SCS. The predicted velocity at L1 using HUX extrapolation is presented in panel (C) of the same figure (magenta line) along with observed value as black dashed line. We note here that the free parameters needed for empirical relations are kept to be the same as in case of CR2053. Predictions of solar wind flow velocity at L1 from other models 1 (red line), 2 (green line) and 3 (blue line) are also shown in panel (C). Models with WSA extrapolation (models 2, 4a) show a better match with observed values as compared to their counterparts using WS extrapolation (models 1, 3). 
The predicted velocity for this case shows an offset of about 1-2 days in predicting the high speed streams observed on 2009-04-19.
We have also carried out the same analysis using the same set of model parameters (see Eq.~\ref{eq:wsa})for the case of CR 2104 that represents the rising phase of the solar cycle 24. 
We find similar trend even for this case as solar wind velocity estimates from Model 2 and Model 4a that involves WSA empirical relation presents a better match with observations as compared those obtained from Model 1 and Model 3. 
\begin{figure}
\centering
\includegraphics[width=\textwidth]{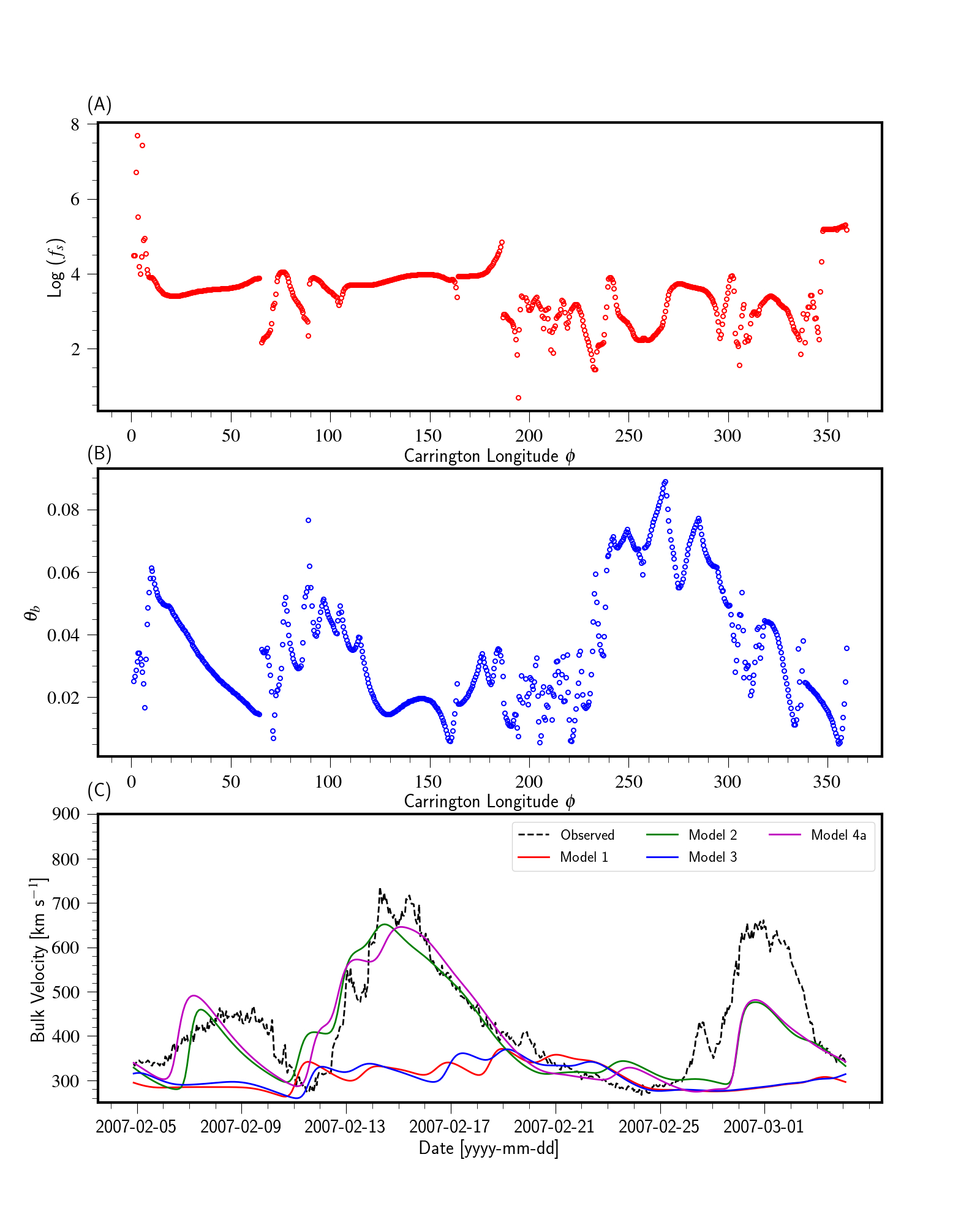}
\caption{Panel (A) and panel (B) show plots of $\theta_b$ and
$f_s$ with respect to the Carrington longitude ($\phi$) for CR 2053. These values are estimated for field lines extrapolated using PFSS+SCS upto a reference sphere of radius 5 R$_{\odot}$. 
Panel (C) shows a comparison of the observed solar wind velocities (taken from OMNI database) and the relative differences between the various outputs of the numerical models at L1 ($\approx 215 R_{\odot}$). A comparison with the physics based model (HD) has been shown later on in the results.}
\label{fig:fs_2053}
\end{figure}

\begin{figure}
\centering
\includegraphics[width=\textwidth]{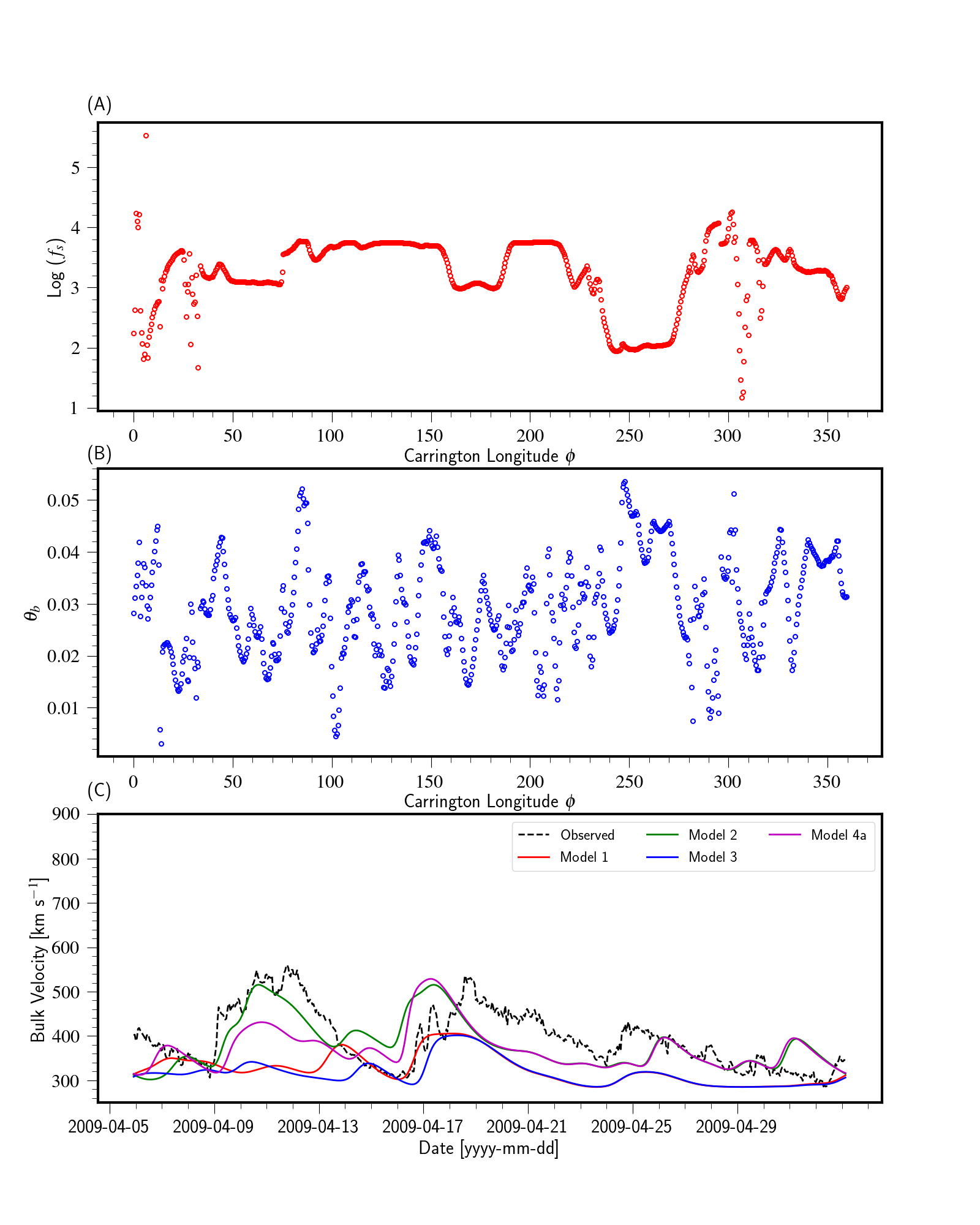}
\caption{Panel (A) and panel (B) show plots of $\theta_b$ and
$f_s$ as a function of the Carrington longitude ($\phi$). These values are estimated for field lines extrapolated using PFSS + SCS up-to the reference sphere of radius 5 R$_{\odot}$ for CR 2082. Panel (C) shows a comparison of the observed solar wind velocities (taken from OMNI database) and the relative differences between the various outputs of the numerical models at L1 ($\approx 215 R_{\odot}$).}
\label{fig:fsthb_vel_2082}
\end{figure}
\subsection{Ensemble Forecasting}
\label{sec:ensemble_fc}
Numerical extrapolation models like HUX are computationally less demanding than HD or 3D MHD models. Such numerical models are thus used to study a large set of initial conditions by a method known as \textit{ensemble forecasting}. Ensemble forecasting has been widely used to constrain terrestrial weather and is an important tool to determine model performance and also helps to set uncertainty bounds to the model output. A solar wind forecast model is considered to be a very uncertain one if the ensemble members produce drastically different results \cite{Reiss:2019}. 
To study the variations introduced in our model output due to an uncertainty in determining source that contribute to solar wind measured at L1, we created an ensemble of latitudes centered around our expected sub-earth latitude at a distance of $\pm  1 ^{\circ}$, $\pm  2.5 ^{\circ}$, $\pm  5 ^{\circ}$, $\pm  10 ^{\circ}$, and $\pm  15 ^{\circ}$. 
The change in sub-earth latitude changes the field lines under consideration which in turn effects the parameters associated with the field lines (e.g,  $\theta_b$, $f_s$) that are used as input to the WSA model. This reflects as a change in the final output velocities at L1. Our ensemble of sub-earth points provides us with a set of 11 different velocities at each Carrington longitude. 
We consider the ensemble median to be the preferred average as an ensemble of sub-earth points are known to produce highly skewed solutions and thus the ensemble mean may not be an accurate representation of the results\cite{Reiss:2019}. 
One should note that the goal of ensemble forecasting in this case is to provide a measure of the uncertainty in the obtained model velocities rather than the ensemble results leading towards a better forecast \citep{Owens:2017, Henley:2017, Reiss:2019}. 
Figure: \ref{fig:ensemble_forecast} shows the forecast obtained from the ensemble of sub-earth latitudes by using Model 4a for all three cases CR 2053, CR 2082 and CR 2104.The dashed line in the figures represent the ensemble median for the respective cases.
We have also plotted the velocity obtained by considering zero uncertainties in the the sub-earth latitudes and the same is plotted as a black solid line.  
The dark and light shades around the median represent the $1\sigma$ and $2\sigma$ variations respectively around the ensemble average. 
A wider spread in the $2\sigma$ values indicates a larger uncertainly in the velocities forecast by the model. 
In the case of CR2053 and CR2082, a large portion of the observed velocity profile lies within the $2\sigma$ error bounds indicating an excellent forecast performance. For the case of CR 2104, the high speed stream is not predicted well with any of our models and some portion of the observed values lie outside of the $2\sigma$ error bounds. This is mainly due to the underestimation of standard deviation from the models as compared to observations. However, the structure is well correlated as evident from high CC (section~\ref{sec:stats_res}).
\subsection{Assessing Interplanetary Magnetic field Polarity}
In addition to predicting the solar wind velocity, we also assess the polarity of the magnetic field lines at L1 assuming a Parker spiral magnetic field configration between Sun and Earth (see \cite{Jian:2015})
We evaluate the corresponding longitude at L1 for each CR longitude at 5 $R_{\odot}$ using the standard streamline equation. The solar wind velocity is assumed to be constant and corresponds to the value obtained using WSA formulation. 
The polarity of radial magnetic field obtained at 5 $R_{\odot}$ from PFSS + SCS extrapolation is then coverted to GSE/GSM co-ordinate frame to compare with the observed polarity of $B_{x}$ component of interplanetary magnetic field at L1 from OMNI database for all three Carrington rotations. The convention of polarity followed at L1 in this case is -1 (outward) and +1 (inward). Figure: \ref{fig:polarity} shows the observed polarity (daily averaged) of inter-planetary magnetic field $B_{x}$ in GSE/GSM co-ordinates and its comparison with that obtained for all three cases CR 2053, CR 2082 and CR 2104 using the field extrapolations and velocity from Model 4a. 
The inter-planetary magnetic field polarity obtained for all three cases show a good agreement with observed daily averaged values. In particular for the case of CR 2104 shown in panel (C) of figure~\ref{fig:polarity}, significant agreement exists  between model prediction and observation. 
\begin{figure}
\centering
\includegraphics[width=\textwidth]{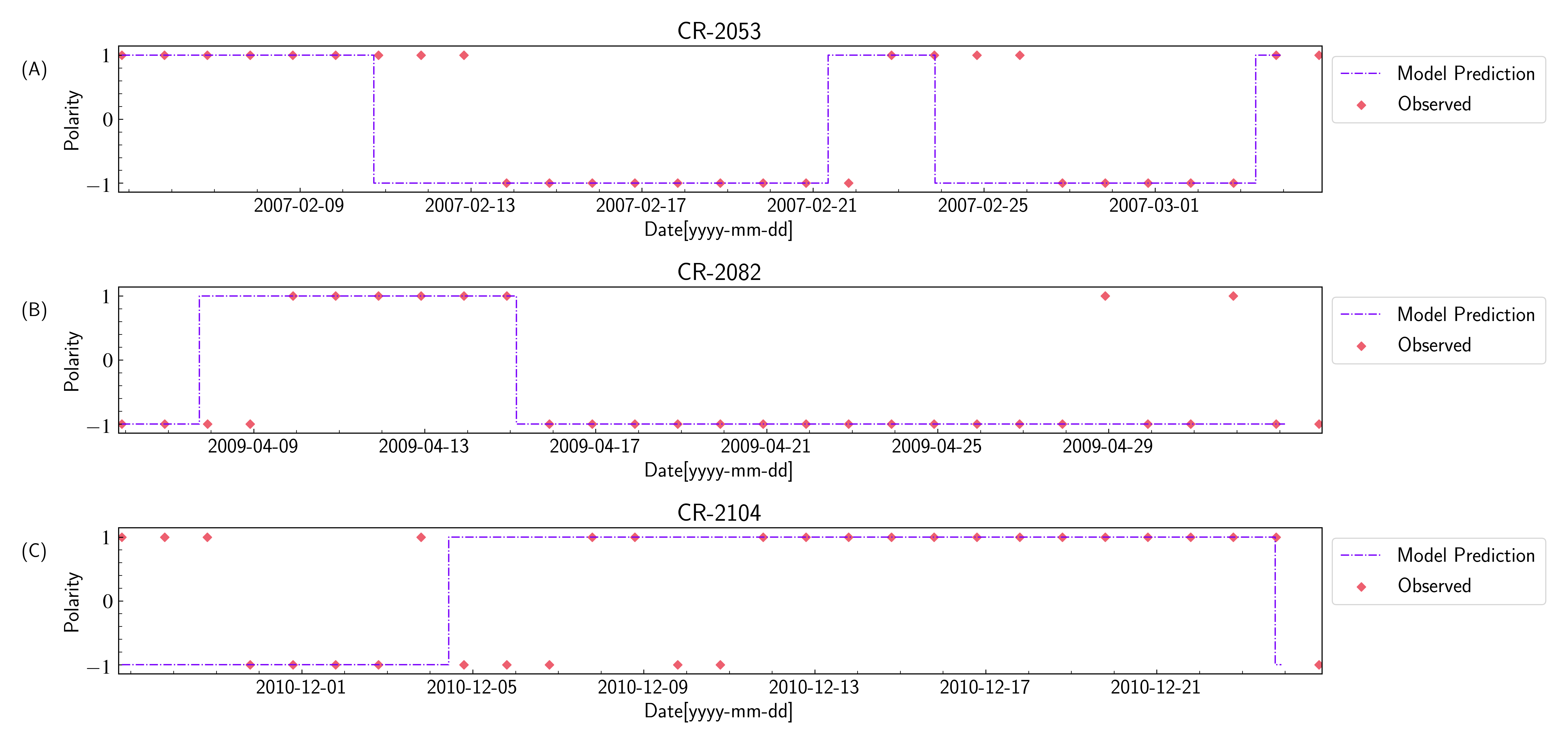}
\caption{A plot showing the magnetic field polarity forecast at L1 for CR 2053 (panel A), CR 2082 (panel B) and CR 2104 (panel C) along with the observed values from OMNI database.
The red diamonds represent the daily average value of polarity of radial component of the interplanetary magnetic field at L1. The purple dashed line are the polarity of the transformed radial component of field lines at L1 in the GSE/GSM co-ordinate frame.}
\label{fig:polarity}
\end{figure}
\subsection{Forecasting from Physics based Modelling}
\label{sec:fc_pluto}
Two dimensional hydrodynamic simulation runs with initial conditions described in section~\ref{sec:pluto_init} provide a time dependent solar wind forecasting framework. 
Plasma is injected in the computational domain from the inner rotating radial boundary at each time step. 
Empirical values of velocities obtained from the WSA model using the sub-earth point field lines are used as input conditions for the hydrodynamic simulations in the inner boundary placed at 5 R$_{\odot}$. 
 
In the initial transient phase, the plasma propagates outwards towards the outer boundary from where it leaves the domain. Subsequent to this transient phase, a steady state solar wind is established in the domain. The steady state radial velocity [X-Y plane] in units of km s$^{-1}$ from the simulation after a one solar rotation time period of 27.3 days is shown in the panel (A) of Figure: \ref{fig:simrun_forecast_2053} for CR2053.
The black dashed circle represents the radial distance corresponding to L1 point ($\sim 1 AU$). Panels (B) and (C) shows the proton temperature in units of MK and number density (cm$^{-3}$) map for the same time step respectively. One can distinctively observe a spiral pattern whereby high velocity streams tend to have higher proton temperature and lower density values than its surrounding low or moderate velocity streams. 
The velocity measured at L1 from simulations during the considered Carrington rotation period is shown as blue solid line in panel D. In comparison, hourly averaged observed values are shown in black dashed lines, while velocity estimates from models 4a and 4b are shown in magneta solid and green dashed lines respectively for comparison. 
The velocity estimate from the simulation runs have more variations as compared to its counterparts from model 4a and 4b. This comparison of models for velocity prediction demonstrates that addition of pressure gradient term and incorporating the energy equation captures the interaction of streams leading to intermittent variable patterns.
Such patterns do not appear for velocities obtained from model 4a and 4b as the kinematic extrapolation disregards physical effects due to presence of velocity shear and pressure gradient terms. Statistical analysis of velocity pattern at L1 obtained from HD simulations indicate a good match with observed values (see Table:~\ref{table_error2053}). Similar variable velocity pattern is also obtained after using the empirical velocity values from CR2082 in the inner boundary (figure not shown). Statistical comparison with other models and observed values is presented in Table:~\ref{table_error2082}.
The advantage of such physics based model is that it accounts for solar wind acceleration and other physical effects which the HUX model lacks. 
Further as the 2D simulations also solves the energy conservation equation with appropriate polytropic equation of state, it has the ability to estimate the proton temperature of the flow at each point in the computational domain. A comparison of proton temperature in MK obtained from simulation run for CR2053 (red solid line) with hourly averaged observed values (black dashed lines) is shown in panel (A) of Figure:~\ref{fig:simrun_temp_2053}. The statistical measures assessing the forecasting performance are mentioned in the panel. 
The curves show a considerable match with a correlation coefficient 0.53 and a RMSE of 0.08 MK. For the case of CR2082 shown in panel (B) of the same figure, the correlation coefficient is 0.32 and RMSE of 0.04 MK. The estimates of standard deviation from modelled and observed data are comparable for the case of CR2082 as well. 

\subsection{Quantifying Predictive Performance using Statistical Analysis}
\label{sec:stats_res}
Statistical measures (see section~\ref{sec:stats}) quantifying forecast performance for CR 2053, CR 2082 and CR 2104 are presented in Table:~\ref{table_error2053}, Table:~\ref{table_error2082} and Table:~\ref{table_error2104} respectively.
The standard deviation of the observed time series for CR2053 is 119.4\,kms$^{-1}$. Model 4c was able to reproduce this standard deviation quite satisfactorily (123.6\,kms$^{-1}$). All models for CR 2053 involving the WS model failed to produce reasonable agreement with the observations, which can be inferred from the extremely low value of the correlation coefficient ($\sim$ 0). We believe that this is due to the proximity of the coronal holes to the equator, whereby open field lines emerging from these holes drive high speed streams. The WS empirical model disregards the size of coronal holes in its empirical relation and fails to capture this effect. The presence of a significant number of coronal holes on the magnetogram that are close to the sub-earth latitudes thus necessitates the use of the WSA model for velocity mapping on the inner boundary. CR2082 on the other hand has a significantly lesser number of coronal holes \cite{Petrie:2013}. And thus even the models involving the WS velocity mapping show acceptable CC values (0.42-0.54). The highest CC for CR 2053 is produced by model 4b (CC= 0.81) which represents the ensemble median of the sub-earth latitude variations, however the same cannot be said for CR 2082 as all the other models produce a greater CC and lower error values than that of the ensemble median. This reinforces the statement that the ensemble average does not always produce a more accurate representation of the velocity profile.

A graphical representation and analysis of various forecasting models implemented in this work are represented in a Taylor diagram \cite{Taylor:2001}. 
The Taylor plots for CR 2053, CR 2082 and CR 2104 are given in Figure:~\ref{fig:taylor_diags}
The radial distance from the origin represents the standard deviation, the azimuthal coordinate represent the correlation coefficient (CC). In addition to this, one more set of arcs are drawn, centered at the reference (observed) standard deviation on the X axis that represent the RMS error for the models. In general a solar wind model is considered to be a good model if it has a high correlation coefficient(CC), low RMS error, and having a similar standard deviation to the observed time series of velocities \cite{Reiss:2016} and in a Taylor plot it would lie close to the smallest circle on the X axis. 
For the case of CR 2053, Model 1 and Model 3 lie very close to the vertical axis indicating very low CC value and thus are considered to have poor forecast performance. Model 4b has the highest CC value and comparatively low errors, but it lies relatively far from the observed standard deviation curve. The physics based models,  Model 4c is seen to have standard deviation profiles well in agreement with the observed values along with a good CC. We thus assert that Model 4b and 4c are the best representations of the velocity profiles for CR 2053.
For the case of CR 2082, Model 4c and Model 2 has a standard deviation closest to that of the observed value of $\sim$ 60\,kms$^{-1}$ and Model 2 has the lowest error values (in terms of RMSE, MAE and MAPE). The CC of model 3 is one of the highest in the models considered followed closely by Model 2. Model 3 however has larger values of all the errors considered showing that even though its correlated well with the observed values, the model is prone to errors which would  result in inaccurate output solar wind velocity profiles. Model 4a , 4b and 4c have similar but low values of MAE,  MAPE and RMSE, and CC lying in the range of ($\sim$ 0.34-0.46). Based on the above analysis for the assumed set of WSA parameters, Model 2 performs better for CR2082 closely followed by models 4c and 4a.
For the case of CR 2104, the predicted standard deviation is less as compared to the observed values for all the models.  The predictions from model 1 and 3 have low CC value and RMSE around 120 km/s and indicating a poor forecast performance. 
The CC of models 2, 4a, 4b and 4c is as high as 0.8 and the RMSE for these models ranges between 80-90 kms/s. This suggests these models have a better forecast performance as was the case in CR 2053. 
\begin{figure}
\centering
\includegraphics[width=\textwidth]{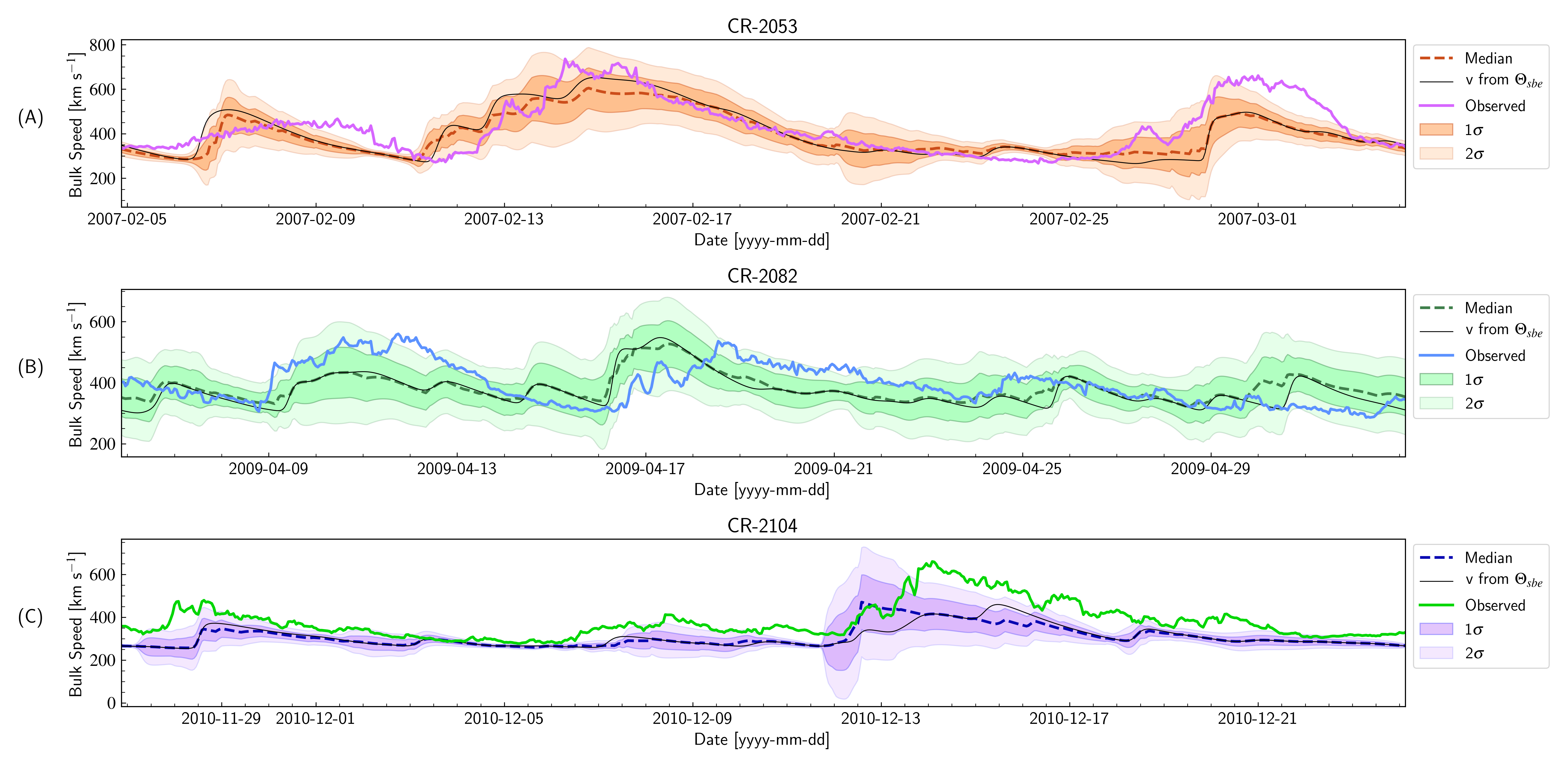}
\caption{A plot of the ensemble forecast for CR 2053 (panel A), CR 2082 (panel B) and CR 2104 (panel C) along with the observed velocities. The black solid line in each plot represents the velocities obtained without incorporating the uncertainty. The dashed line in each of the plots represent the respective ensemble median values and the darker and lighter shades around the median values represent the $1\sigma$ and $2\sigma$ error expected during the forecast respectively.}
\label{fig:ensemble_forecast}
\end{figure}
\begin{table}
\setlength{\tabcolsep}{10pt}
\renewcommand{\arraystretch}{1.5}
\caption{\label{table_error2053} Performance of various forecast models for CR 2053}
\begin{center}
    \begin{tabular}{l c c c c c c }
       \hline\hline
        $\sigma_{obs}$= 119.4 $kms^{-1}$\vline& Model 1 & Model 2 & Model 3 & Model 4a & Model 4b & Model 4c \\ \hline\hline
        $\sigma$ & 31.1   & 105.8  & 30.1   & 109.9   & 90.8    &123.6    \\ \hline
        MAE      & 131.1  & 60.8   & 125.0  & 59.2    & 57.8    & 66.9    \\ \hline
        MAPE     & 26.5   & 14.2   & 24.8   & 13.5    & 12.9    & 14.9    \\ \hline
        RMSE     & 171.8  & 79.4   & 169.5  & 77.9    & 75.5    & 88.8    \\ \hline
        CC       & 0.00   & 0.79   & 0.04   & 0.79    & 0.81    & 0.73     \\\hline
        \hline
\end{tabular}
\end{center}
\end{table}
\begin{table}
\setlength{\tabcolsep}{10pt}
\renewcommand{\arraystretch}{1.5}
\caption{\label{table_error2082} Performance of various forecast models for CR 2082}
\begin{center}
    \begin{tabular}{l c c c c c c}
       \hline\hline
        $\sigma_{obs}$= 66.0$kms^{-1}$\vline  & Model 1 & Model 2 & Model 3 & Model 4a & Model 4b & Model 4c \\ \hline\hline
        $\sigma$ & 37.1   & 60.8   & 32.2   & 51.7    & 42.9    & 59.4    \\ \hline
        MAE      & 73.0   & 52.1   & 79.3   & 56.4    & 54.6    & 52.8    \\ \hline
        MAPE     & 17.1   & 13.3   & 18.7   & 13.8    & 13.6    & 13.3    \\ \hline
        RMSE     & 92.1   & 63.3   & 95.8   & 69.3    & 66.5    & 65.3    \\ \hline
        CC       & 0.42   & 0.52   & 0.54   & 0.39    & 0.34    & 0.46    \\ \hline
        \hline
\end{tabular}
\end{center}
\end{table}
\begin{table}
\setlength{\tabcolsep}{10pt}
\renewcommand{\arraystretch}{1.5}
\caption{\label{table_error2104} Performance of various forecast models for CR 2104}
\begin{center}
    \begin{tabular}{l c c c c c c}
       \hline\hline
        $\sigma_{obs}$= 82.2 $kms^{-1}$\vline  & Model 1 & Model 2 & Model 3 & Model 4a & Model 4b & Model 4c \\ \hline\hline
        $\sigma$ & 26.0   & 57.0   & 27.9   & 47.4    & 47.6    & 49.1    \\ \hline
        MAE      & 93.6   & 68.7   & 94.3   & 72.3    & 74.0    & 65.7    \\ \hline
        MAPE     & 21.9   & 17.2   & 22.4   & 17.8    & 18.1    & 16.0    \\ \hline
        RMSE    & 123.6   & 79.2   & 118.6  & 87.2    & 90.0    & 81.1    \\ \hline
        CC       & 0.00   & 0.88   & 0.35   & 0.84    & 0.79    & 0.85    \\ \hline
        \hline
\end{tabular}
\end{center}
\end{table}

\section{Discussions and Summary}
\label{sec:discuss}
\begin{figure}
\centering
\includegraphics[width=\textwidth]{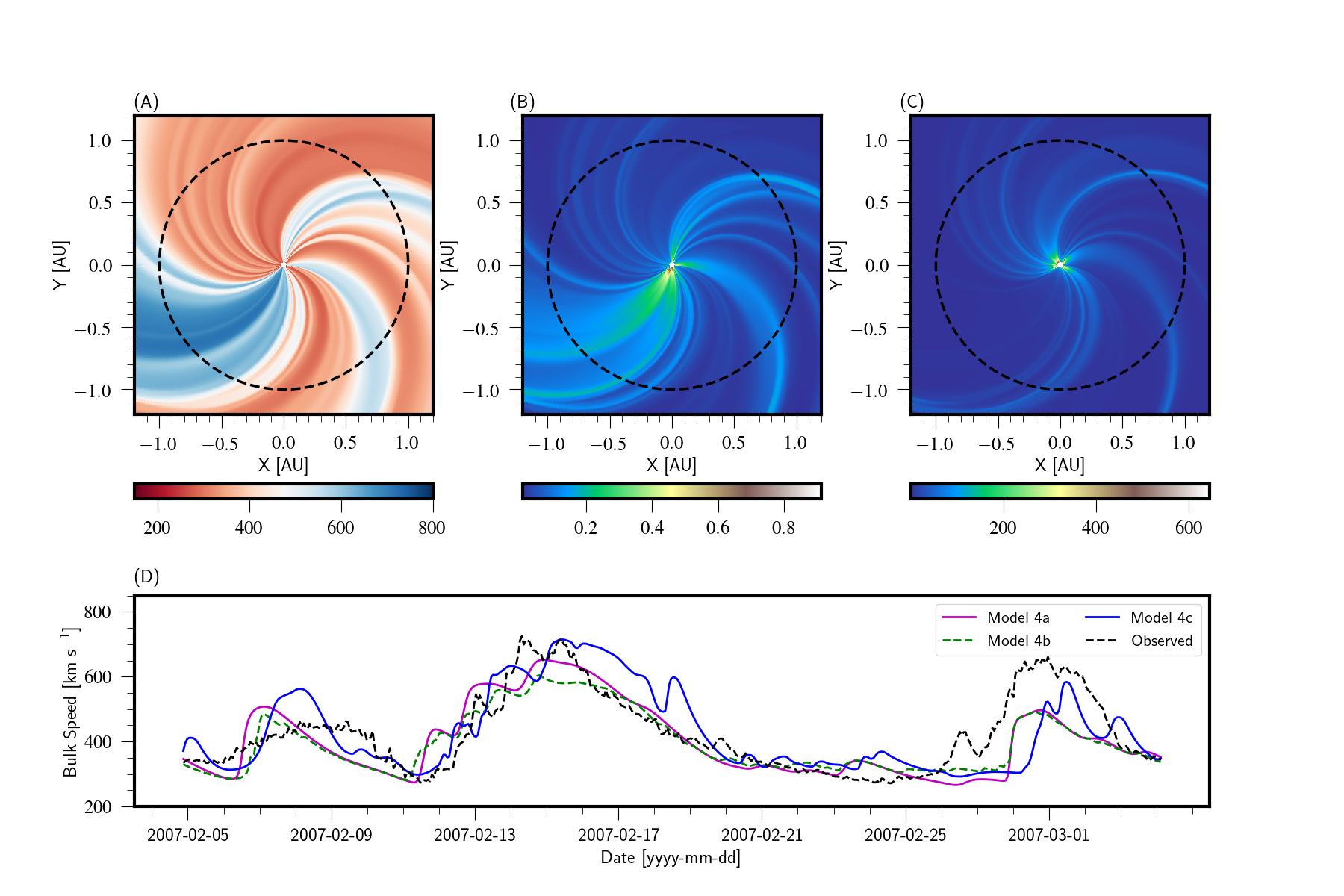}
\caption{Quantities of the solar wind obtained from the physics based model (A) Radial velocity (B) Proton temperature in MK and (C) number density in cm$^{-3}$, the black dashed line in each of these panels represents the L1 position at 1 AU. Panel (D) shows comparison of variation in bulk speed in kms$^{-1}$ at L1 point for CR 2053 for Models 4a, 4b and 4c with hourly averaged observed values. The X-axis in panel (D) represents the time in yyyy-mm-dd format.}
\label{fig:simrun_forecast_2053}
\end{figure}
In this pilot study, we have developed a python module towards constructing a robust framework for accurate predictions of a steady state background solar wind model using various empirical and extrapolation formulations. 
This framework is also integrated with physics based modelling using PLUTO code. 
The entire workflow is a combination of \textit{(a)} extrapolating the magnetic fields to the outer coronal domain using magnetic models such as PFSS and SCS, \textit{(b)} mapping the velocities in the outer coronal domain using models such as WS and WSA and \textit{(c)} extrapolating the velocities to L1 using extrapolation techniques such as HUX as well as hydrodynamic propagation of the velocities using PLUTO. We have studied various combinations of the coronal magnetic models as well as velocity extrapolation models in view of three different Carrington rotations, CR 2053, CR 2082 and CR 2104. We were successfully able to generate velocity profiles well in agreement with the observed values.  
    
Even though the the models involving the WS velocity mapping performed relatively well for the case of CR 2082 when compared to the equivalent models of the other CRs in this study,  it performed very poorly when applied to CR 2053 and CR 2104. The WSA model on the other hand has shown consistently superior performance for all the CRs with lesser error measures in these cases than the WS model. The WS model also failed to capture the contrast between the slow and fast solar wind streams and thus produced velocity profiles with significantly lesser standard deviation than the observed profile. We infer from this that the WSA model is superior to the WS model for the cases and parameters considered in this work.   
    
In general, the models with a combination of PFSS + SCS for magnetic field extrapolation paired with the WSA model for velocity mapping (models 4a,  4b and 4c) had the best performance in all the cases, CR 2053, CR 2082 as well as CR 2104. This can be seen as near identical standard deviations of the output velocity pattern in Model 4a and Model 4c paired with a very high correlation coefficient ($\sim$ 0.73-0.81) and relatively low errors for the case of CR 2053 (Table: \ref{table_error2053}) and CR 2104 (Table: \ref{table_error2104}). For CR 2082, Model 4a and Model 4c had standard deviations which are close to that of the observed profile. We note that Model 3 performed quite well in case of CR 2082 with a high CC ($\sim$ 0.53) but with higher errors than the other models. However, it showed a lack of consistency in performance with very poor results in case of CR 2053. We observe that Model 2 also performed well for all three CRs which can be seen in its high CC values and low errors. 

Model 4c that employs PFSS+SCS and WSA velocity empirical relation with physics based extrapolation, gives us an additional advantage of being able to determine the thermal properties of the solar wind which also showed good correlation with the observed values (see Figure: ~\ref{fig:simrun_temp_2053}). This model also naturally takes into account the acceleration of the solar wind during its propagation. The standard deviation of the Model 4c velocity profile also had good agreement with the observations with high CC values (0.73 for CR 2053, 0.46 for CR 2082 and 0.85 for CR 2104). The Model 4c thus has its own pros and cons. The pros being the ability to determine additional solar wind properties than what HUX or HUXt can offer, and the cons being that it takes significantly more computational resources.
    
Ensemble forecasting helps provide a rather clear quantification of the uncertainty associated with the predictions. We have chosen a total of 11 realisations covering a latitude range of $\pm 15^{\circ}$ about the expected sub-earth latitude. We can assert that the variation helped us accurately capture the uncertainty in the predictions as a bulk of the observed velocity curve was entirely within the $2\sigma$ bounds of the model output for CR 2053 as well as CR 2082 (Figure: \ref{fig:ensemble_forecast}). Underestimation in the solar wind velocity is seen in case of CR 2104 particularly for the high speeds streams. 
We also note that the ensemble median did not necessarily provide a better estimate of the model velocity and in general, had sub-par performance when compared to some of the other models (Model 4a and Model 4c). 
The radial magnetic field polarity obtained at 5 R$_{\odot}$ has also been extrapolated to L1 assuming a Parker spiral field profile for Model 4a. A good agreement between the observed and predicted field polarity is seen in all three cases with CR 2082 showing misses only for three data points obtained from observations.

Even though ensemble forecasting can provide a rather clear statistical uncertainty prediction of the model, it has been seen that various other uncertainties can creep in that are solely dependent on the choice of the input magnetograms. Riley et.al \cite{riley_diff_magneto} highlighted that magnetograms from various observatories show significant differences in magnetic field measurements. In-spite of the fact that one can quantify the conversion factors between various independent datasets, there is no particular observatory that can produce a \textit{ground truth} dataset for the input magnetograms. 
Additionally, one can improve the WSA model performance using  time-dependent flux transport model based magnetograms \citep[e.g.,][]{Schrijver:2003, Arge:2010, flux_transport}.
Quantifying the effects of input magnetograms on model performance is beyond the scope of this paper and shall be addressed in a future study.
\begin{figure}
\centering
\includegraphics[width=0.32\textwidth]{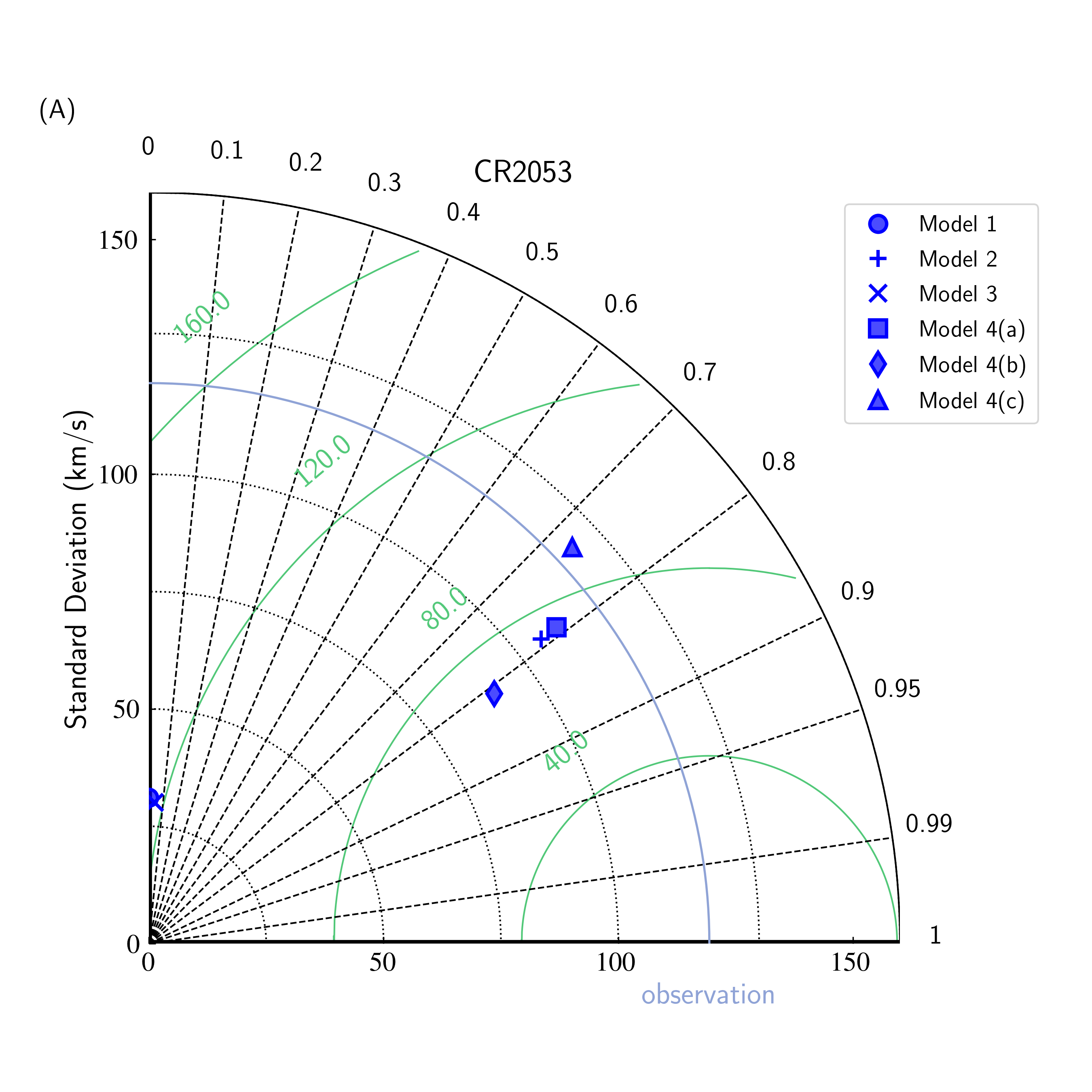}
\includegraphics[width=0.32\textwidth]{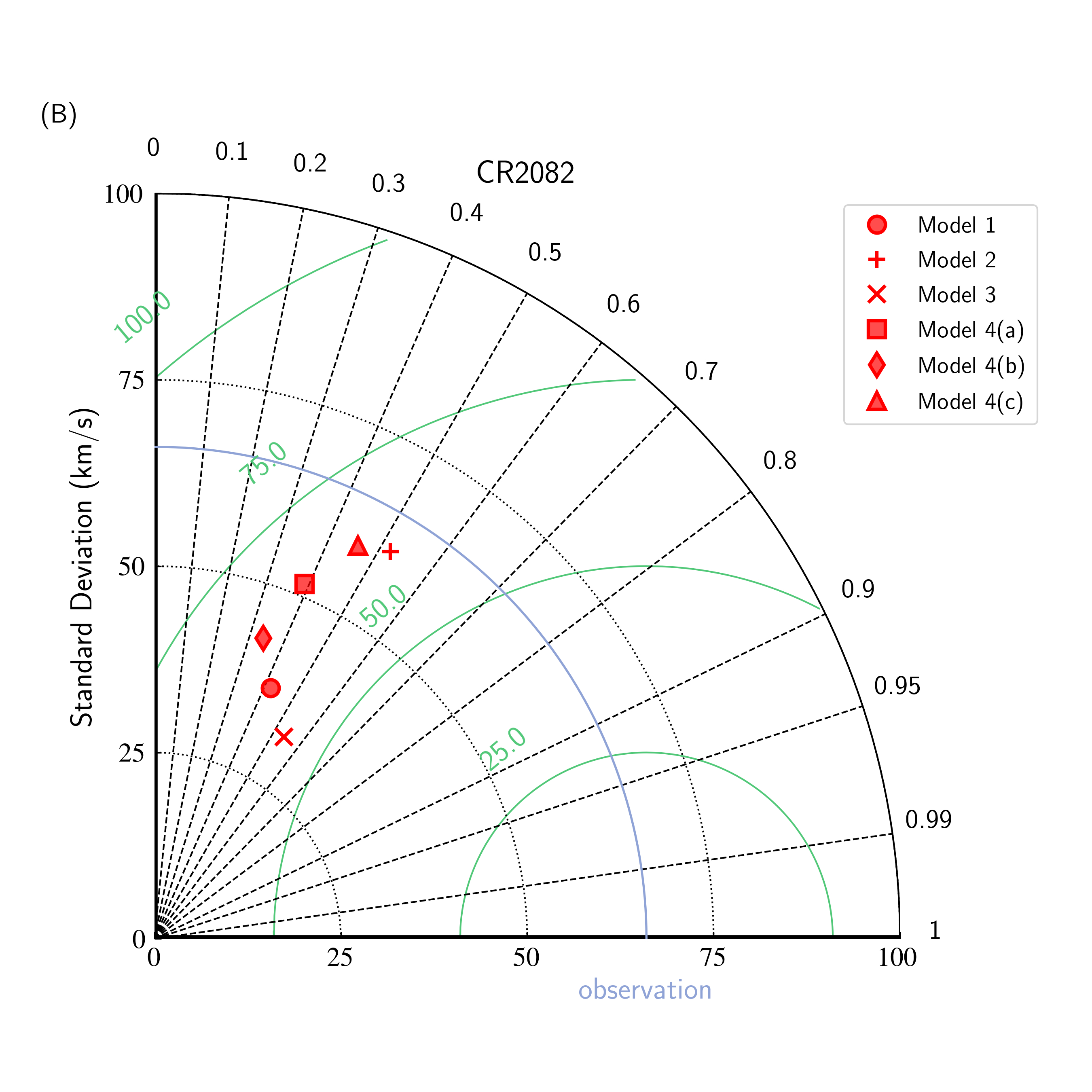}
\includegraphics[width=0.32\textwidth]{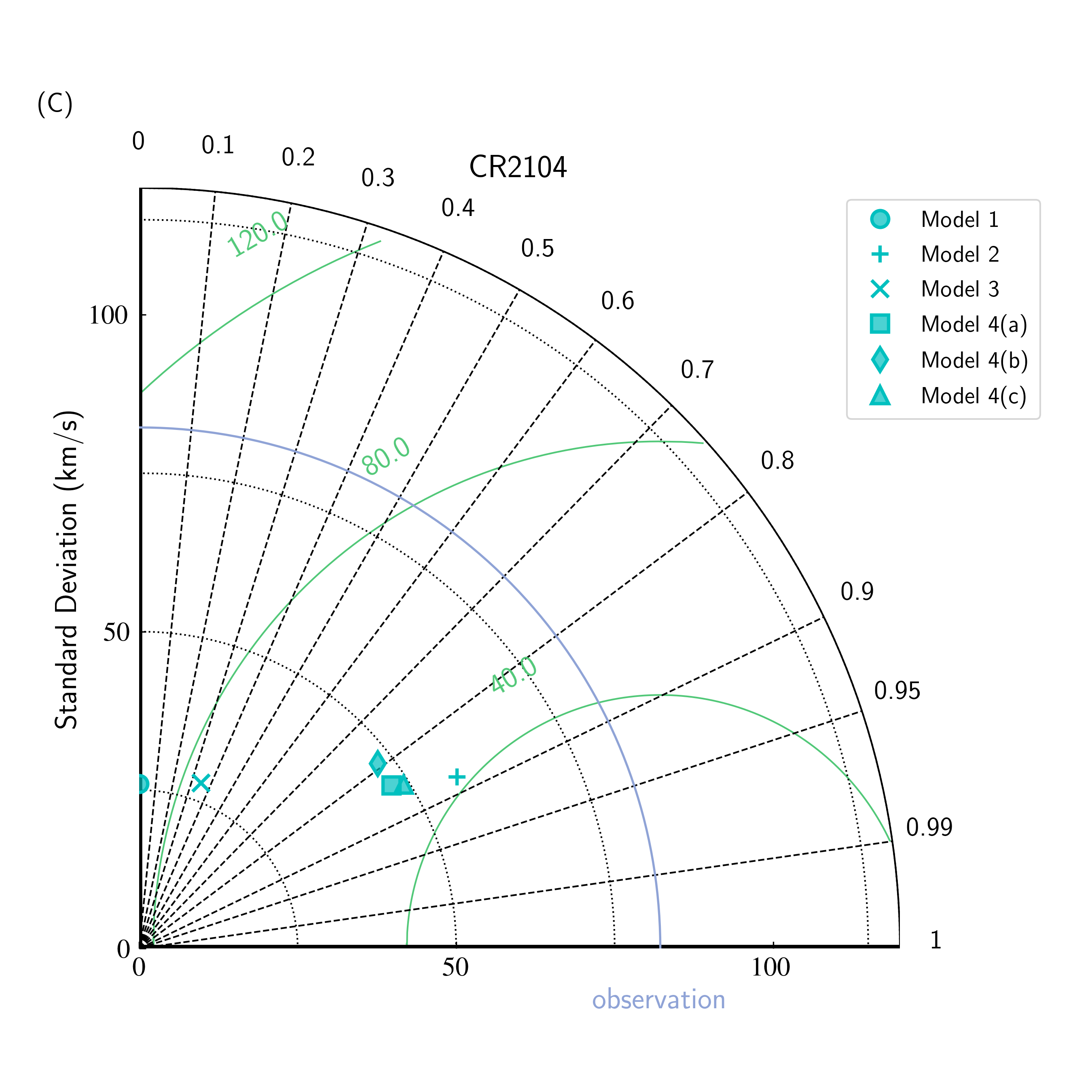}
\caption{Taylor Plot for CR 2053 (panel A), CR 2082 (panel B) and CR 2104 (panel C) summarizing the performance of various models in each case. The blue curve concentric to the standard deviation curves represents the standard deviation of the observed time series. The green curves with centers on the X axis serve as ticks for the RMSE.}
\label{fig:taylor_diags}
\end{figure}    
The statistical parameters represented in the Taylor diagrams indicate the accuracy of our current forecasting framework. Typically, for an accurate forecasting model, low RMSE, high CC and similar standard deviation with observed values are expected. The forecasting presented in this paper uses the same set of parameters (except $\theta_b$ and $f_s$) for empirical velocity estimate of all the Carrington rotations. We find that these parameters give a more accurate forecast of the bulk solar wind speeds at L1 for the case of CR 2053. Even though the same parameters result in good forecast for CR 2082 and CR 2104 as well, we believe that the performance may be improved with a changes in the free parameters and radial distance of reference sphere \cite{Wu:2020aa}.

We would like to point out that the free parameters used for the WS and WSA empirical relations cannot be universally applied for all CRs and must be individually tuned for a particular case scenario. This is evident from our use of the same set of free parameters for all the CRs which results in significantly different performances.
Additionally, the statistical estimates have indicated that in case of CR 2082, the model with PFSS+WSA have performed slightly better than its counterpart including SCS. 
In this particular case, this could perhaps be related to over-estimation of magnetic field lines contributing to solar wind speeds at L1 by SCS. However, through the ensemble modelling we have demonstrated the estimation from Model 4a is within the 2$\sigma$ bounds. Improving the accuracy and performance of SCS extrapolation will be considered in our further studies.
\begin{figure}[H]
\centering
\includegraphics[width=0.95\textwidth]{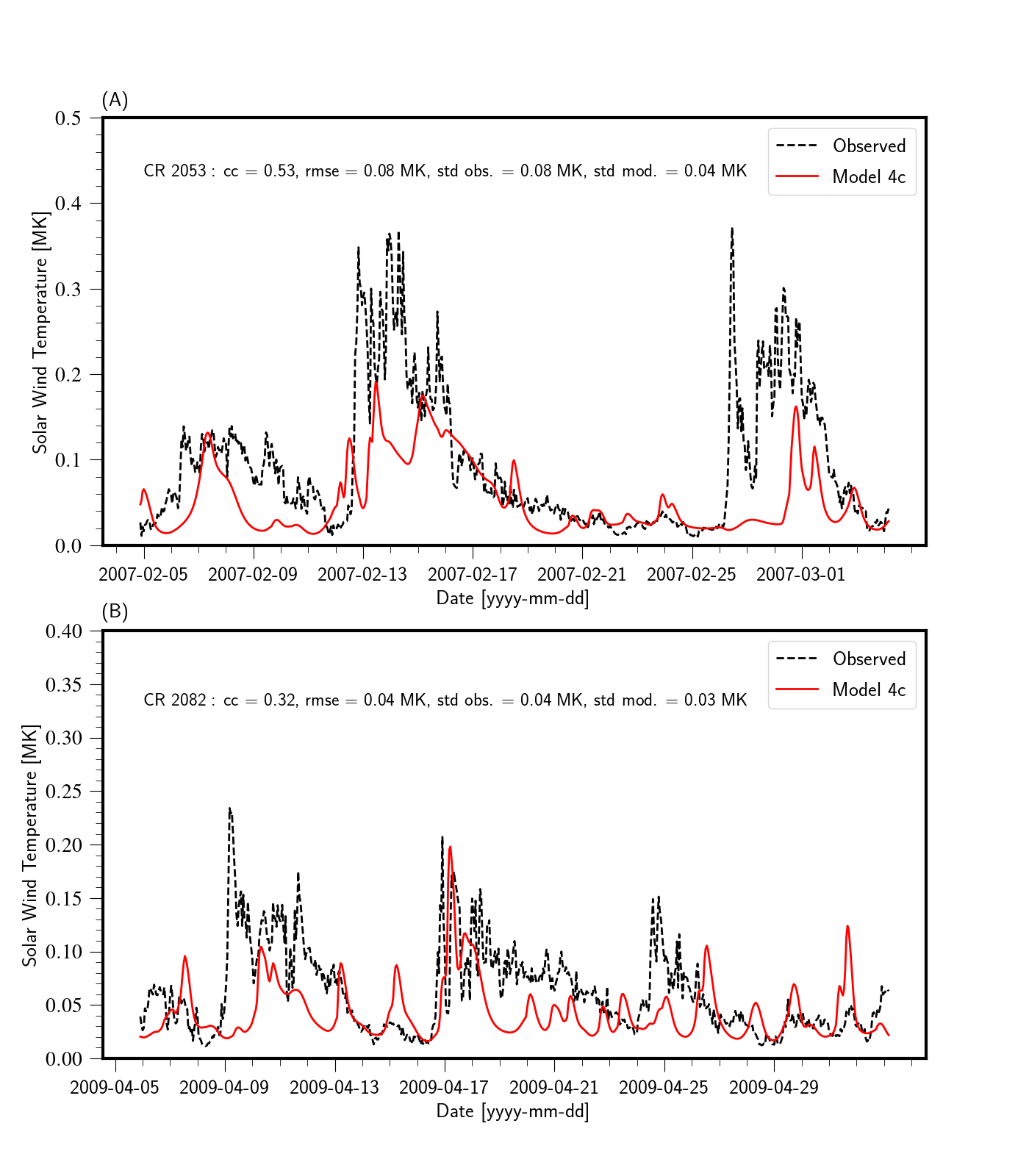}
\caption{(A) Comparison of the proton temperature estimated from the model 4c (red solid line) at Lagrangian point L1 with hourly averaged observed values (black dashed line) obtained from OMNI database for CR2053. The statistical analysis of the forecast performance are mentioned in the panel. (B) Comparison ofproton temperature estimates from model 4c and observations for the case of CR2082. The color scheme is same as the panel above.}
\label{fig:simrun_temp_2053}
\end{figure}
The caveat of hydrodynamic models (Model 4c) is that it solves the equations on a plane and does not contain the magnetic information of the sun-earth system on which state-of-the-art 3D MHD model are based. However, even without the magnetic field information and  having a 2D geometry, the HD models capture many important aspects of the solar wind. 
Model 4c facilitate better physical insight in the behaviour of solar wind than the HUX models (4a and 4b). Model 4c (HD) can thus be seen as a halfway point between HUX and full MHD models.

In summary, we have successfully produced velocity maps for the cases considered and also matched our results with additional observable e.g, proton temperature. This pilot study is the first step in developing an indigenous space weather framework which is an absolute need of hour as Indian Space Research Organisation (ISRO) is coming up with Aditya-L1 to study the properties of Sun at L1 (\url{https://www.isro.gov.in/aditya-l1-first-indian-mission-to-study-sun}). A full-fledged 3D-MHD run of solar wind using PLUTO code will be carried out in our subsequent paper for predicting other observable quantities like the number density and IMF magnetic field. Further, including the propagation of CMEs in such a realistic solar wind background would be carried out with an aim to study its impact on space weather.

\bibliographystyle{frontiersinSCNS_ENG_HUMS}
\bibliography{mainbody.bib}

\end{nolinenumbers}
\end{document}